\documentstyle[aps,prb,multicol]{revtex} 
\input epsf

\newcommand{\dir}{Figs}
\newcommand{\fig}[4]
{
    \noindent
    \unitlength=1mm
    \begin{picture}(#2,#3)
    \put(10,0){\leavevmode \epsfxsize=#2mm \epsffile{\dir/#1}}
    \put(5,45){#4}
    \end{picture}
    \noindent 
}
\newcommand{\rr}{ {\bf r} }
\newcommand{\ru}{ {\bf \hat{r}} }

\newcommand{\nn}{ {\bf n} }
\newcommand{\uu}{ {\bf u} }
\newcommand{\kk}{ {\bf k} }
\newcommand{\QQ}{ {\bf Q} }

\newcommand{\II}{ {\bf I} }
\begin{document} 

\newcommand{\CCshift}
{
\caption{
 Expansion coefficient $h_{212-100}$ of the total correlation function
 vs. $1/r$ for systems of different size at density $\rho=0.3 \sigma_0^{-3}$.
 Before performing a Hankel transform of this curve, the tails are
 fitted with a function $a + b/r$, and the whole curve is shifted by $a$.
 Inset shows the shifted data. The curves for the two larger systems
 are then almost on top of each other.
}
\label{fig:shift}
}

\newcommand{\CCgubbins}
{
\caption{
 Illustration of the Gubbins equation (\ref{eq:gubbins2}) for $l=2$ and 4
 at the density $\rho=0.3 \sigma_0^{-3}$. Open symbols show right 
 hand side of eqn. (\ref{eq:gubbins2}) vs. $k^2$, 
 closed symbols the value at $k \to 0 $ obtained from the left hand side.
 The system size was $N=8000$.
}
\label{fig:gubbins}
}

\newcommand{\CCaver}
{
\caption{
  Orientational average of the pair distribution function $\rho^{(2)}$ (a), 
  the total correlation function $h$ (b),
  and the direct correlation function $c$ (c)
  vs. distance $r$ for different densities. 
  Data for different system sizes ($\rho = 0.3 \sigma^{-3}$)
  lie on top of each other.
}
\label{fig:aver}
}

\newcommand{\CCelastic}
{
\caption{
  Expansion coefficient with $l_i=l=2$ and $m_1=-m_2=1$ of
  the pair distribution function $\rho^{(2)}$ (a),
  the total correlation function $h$ (b),
  and the direct correlation function $c$ (c)
  vs. distance $r$
  for the density $\rho=0.24 \sigma_0^{-3}$ (long dashed line)
  and $\rho=0.3 \sigma_0^{-3}$ (solid line: system size $N=8000$,
  dotted line: $N=4000$, short dashed line: $N=1000$).
  Dashed dotted line in the plot for $h$ indicates extrapolation 
  with a $1/r$ behavior. Inset shows data for $h$ as a function 
  of $1/r$.
}
\label{fig:elastic}
}

\newcommand{\CCorient}
{
\caption{
  Expansion coefficient with $l_i=2$ and $l=m_i=m=0$ of
  the pair distribution function $\rho^{(2)}$ (a),
  the total correlation function $h$ (b),
  and the direct correlation function $c$ (c)
  vs. distance $r$
  for different densities.
  Dot dashed line in the plot for $\rho^{(2)}$ indicates infinite 
  range limit for $\rho=0.287 \sigma_0^{-3}$.
  Inset of (b) shows blowup of data for large distances.
}
\label{fig:orient}
}

\newcommand{\CCframes}
{
\caption{
  Illustration of the coordinate frames. Thick solid lines
  show coordinate axes of director frame, where the $z$ axis
  points in the direction of the director $\nn$. Thick dashed
  lines indicate axes of the molecular frame, where the
  $z$-axis points along the vector $\rr_{12}$ which connects 
  the two particles. The $x$ axis of the molecular frame lies
  in the plane which is spanned by $\rr_{12}$ and $\nn$.
} 
\label{fig:frames}
}

\newcommand{\CCnibreak}
{
\caption{
  Expansion coefficient with $l_i=l=2$ and $m_i=m=0$ 
  of the direct correlation function in the molecular
  frame vs. $r$ for different densities. 
  In the isotropic phase, this coefficient must
  vanish for symmetry reasons. At $\rho = 0.283 \sigma_0^{-3}$,
  a small nonzero remnants are observed due to finite size
  effects.
} 
\label{fig:nibreak}
}

\newcommand{\CCnisym}
{
\caption{
  Expansion coefficient with $l_i=2$ and $l=m_i=m=0$ 
  of the direct correlation function in the molecular
  frame vs. $r$ for different densities. 
} 
\label{fig:nisym}
}

\newcommand{\CCpy}
{
\caption{
  Expansion coefficient with $l_i=2$ and $l=m_i=m=0$ 
  of the direct correlation function in the molecular
  frame vs. distance $r$ in Percus-Yevick approximation 
  and in simulations at the density 
  $\rho=0.24 \sigma_0^{-3}$ (a)
  and $\rho=0.283 \sigma_0^{-3}$ (b).
  Simulation data are the same as in fig. \ref{fig:nisym}.
  (same data as figure \ref{fig:nisym}).
  Dashed line shows Percus-Yevick prediction for 
  $\rho=0.3 \sigma_0^{-3}$ for comparison.
} 
\label{fig:py}
}

%
%
\title{Local structure in nematic and isotropic liquid crystals}

\author{Nguyen Hoang Phuong and Friederike Schmid}

\address{Fakult\"at f\"ur Physik, Universit\"at Bielefeld, 
         33615 Bielefeld, Germany}

\setcounter{page}{1}
\maketitle 

\begin{abstract}
\noindent
By computer simulations of systems of ellipsoids, we study the influence 
of the isotropic/nematic phase transition on the direct correlation 
functions (DCF) in anisotropic fluids. The DCF is determined from the pair 
distribution function by solving the full Ornstein-Zernike equation, without 
any approximations. Using a suitable molecular-fixed reference frame, 
we can distinguish between two qualitatively different contributions to 
the DCF: One which preserves rotational invariance, and one which breaks 
it and vanishes in the isotropic phase. We find that the
symmetry preserving contribution is barely affected by the phase
transition. However, symmetry breaking contributions emerge 
in the nematic phase and may become quite substantial. Thus the DCF 
in a nematic fluid is not rotationally invariant.
In the isotropic fluid, the DCF is in good agreement with the 
prediction of the Percus-Yevick theory.

\end{abstract}

%
%

\section{Introduction} 

\label{sec:intro}
\begin{multicols}{2}

The pair direct correlation function (DCF) is a central quantity in the 
theory of liquids. Through the Ornstein-Zernike relation~\cite{hansen}, 
it is related to the pair distribution function, 
which is accessible experimentally. Compared to the latter, it often
has a much simpler structure, because the Ornstein-Zernike relation 
eliminates to a large extent the contributions of third particles to 
the pair correlations. Moreover, the DCF plays a key role in density 
functional theories, since it is the second functional derivative of the 
excess free energy with respect to the local density~\cite{hansen,evans}. 
Much effort has therefore been devoted to the investigation of DCFs 
in molecular fluids. 

Most studies have considered isotropic fluids: Early integral equation 
approaches are due to Wulf~\cite{wulf1}, and Chen and Steele~\cite{chen}. 
Chen and Steele generalized the Ornstein-Zernike equation to the case 
of particles with orientation dependent interactions~\cite{chen,gubbins} 
and calculated the structure of diatomic fluids using a 
Percus-Yevick~\cite{percus} closure. Percus-Yevick and Hypernetted 
Chain~\cite{hansen} theories for isotropic fluids with orientation 
dependent interactions have also been introduced by Pynn~\cite{pynn}, 
Fries and Patek~\cite{fries}, and Singh and coworkers~\cite{singh1}. 
Several authors have calculated DCFs in isotropic liquid crystals using 
theories of this kind~\cite{pynn,perera,pospisil,singh2,letz}. 
The resulting pair correlation functions were generally in reasonable 
agreement with available simulation data. 

Alternative approaches have been pursued, e. g., by Wulf~\cite{wulf1,wulf2}, 
Rickayzen and coworkers~\cite{rickayzen1,rickayzen2,rickayzen3}, and Chamoux
and Perera~\cite{chamoux}. Rickayzen et al choose a geometrically
motivated Ansatz for the form of the DCF and optimize the parameters
such that the Percus-Yevick relation is fulfilled in a given limit.
Chamoux and Perera derive approximations for the DCF from the Rosenfeld 
density functional~\cite{rosenfeld}. For fluids of elongated particles, 
their DCF is similar to that obtained with the Percus-Yevick or 
Hypernetted chain calculations of Perera et al~\cite{perera}.
A number of other authors have suggested simple approximate expressions for 
the DCF in isotropic molecular fluids~\cite{pynn,wulf2,parsons,lee,baus,marko}.
Some of these fit simulation data surprisingly well~\cite{allen1}.
The direct determination of the DCF from computer simulations is
rather tedious. Allen et al~\cite{allen1} have presented a method which 
allows to calculate the DCF from pair correlation data, and applied
it to study the DCF in fluids of ellipsoids~\cite{allen1} and of
Gay-Berne~\cite{gay} particles~\cite{allen2,allen3}.

The DCF has thus been studied quite intensely in isotropic fluids.
In contrast, only few studies have considered the DCF in anisotropic,
e. g., nematic fluids. In a nematic liquid crystal, the particles
remain positionally disordered, but align preferentially along one 
(arbitrary) direction~\cite{degennes,chandrasekhar}. The isotropy of space
is spontaneously broken. Consequently, the pair correlation functions 
lose their rotational invariance. Moreover, soft Goldstone modes
must be present, since the orientational order breaks a continuous
symmetry: The structure factor diverges in the $\kk \to 0$ limit,
and the pair distribution function exhibits a long range elastic tail.
This implies that the DCF must fulfill certain conditions in the 
$\kk \to 0$ limit, which have been derived by Gubbins~\cite{gubbins},
and later in a different manner by Zhong and Petschek~\cite{zhong1}.

Workman and Fixman~\cite{workman} have formulated the general form
of the Ornstein-Zernike equation for the case of anisotropic fluids.
They also suggested a Percus-Yevick closure which is based on a density 
functional about an isotropic reference state, but can be applied to
both isotropic and anisotropic liquids. In fact, the Percus-Yevick 
and the Hypernetted chain closure lend themselves to a rather
straightforward generalization for anisotropic fluids~\cite{caillol}.
Caillol and coworkers~\cite{caillol} have used these closures to 
calculate the structure of perfectly aligned fluids. 
Zhong and Petschek~\cite{zhong1} have analyzed the diagrammatic 
expansion of the DCF in this approach, and proved that at least the
Percus-Yevick closure must fail to reproduce the soft Goldstone
modes in the general case of spontaneous partial order.
In order to fix the problem without resorting to a reference state, 
they proposed a modified Percus-Yevick closure which premises that 
the DCF is rotationally invariant. Given that the DCF is vaguely related 
to an ``effective interaction potential'' between 
particles~\cite{percus,zhong1}, this assumption seems plausible.
Nevertheless, a clearcut test of the hypothesis is clearly desirable.

In practice, theoretical studies of DCFs in partially ordered nematic
fluids have mostly been concerned with systems with separable 
interactions~\cite{zhong2,holovko}, where the interparticle potential 
depends on the spatial separation and the orientation of particles 
independently. Very little is still known on the DCF in general 
anisotropic fluids. Simulation studies have been performed by 
Stelzer et al~\cite{stelzer} (nematic Gay-Berne fluids), and 
by Zakharov and Maliniak~\cite{zakharov} (5CB molecules).
These authors used an ``unoriented nematic approximation'', 
which replaces the pair correlations by their rotational averages.
The simplification has later been questioned by Longa et al~\cite{longa}. 
Indeed, the elastic properties of the nematic fluid, which can be 
calculated from the DCF using the Poniewierski-Stecki equations~\cite{ps}, 
don't seem to be captured very well~\cite{allen4}. Moreover, an analysis which
uses such an approximation is clearly not suited to elicit whether or 
not the DCF depends on the direction of prefered alignment in the nematic fluid.

In a previous paper~\cite{phuong1}, we have presented a method for 
determining the DCF in uniaxial nematic fluids without approximations 
from computer simulations. As a test of the method, we have applied 
it to a nematic fluid of soft ellipsoids and calculated the elastic 
constants from the DCF via the Poniewierski-Stecki equations~\cite{ps}. 
The results were in good agreement with those obtained independently from 
an analysis of order tensor fluctuations, following a procedure by Allen 
et al~\cite{allen4}. Thus we have shown that the method takes into account 
the elastic properties of the fluid in an adequate way.

In the present work, we use our method to analyze the form of the DCF 
in detail for different state points in the nematic and the isotropic
phase. In particular, we discuss the symmetry properties and the range 
of the DCF, the pair distribution function, and the total correlation
function. We show that the DCF is truly short ranged as expected. 
The elastic tails and the near-critical long range fluctuations 
close to the nematic-isotropic transition disappear. On the other hand, 
our data show that the DCF does reflect the broken symmetry of 
the nematic phase, and the assumption of rotational invariance is not 
correct in our system.

The paper is organized as follows: In the next section, we 
define the simulation model, introduce the pair correlation functions
and explain how we perform the data analysis. The results are presented and 
discussed in section \ref{sec:results}. In the isotropic phase, 
we compare them with the prediction of the Percus-Yevick theory. 
Some comments on the Percus-Yevick solution and how it breaks down 
at large densities are added in the appendix. We summarize and conclude 
in section \ref{sec:summary}.

\section{Model and method}

\subsection{simulation model}
\label{sec:model}

We have studied a fluid of soft ellipsoidal particles with repulsive pair 
interactions
\begin{equation}
V_{12}
= \left\{ \begin{array}{lcr}
4 \epsilon_0 \: (X_{12}^{12} - X_{12}^{6}) + \epsilon_0 & : & X_{12}^6 > 
1/2 \\
0 & : & \mbox{otherwise}
\end{array} \right. .
\end{equation}
The function  $X_{12} = \sigma_0/(r_{12}-\sigma_{12}+\sigma_0)$  
depends on the distance $r_{12}$ between particles $1$ and $2$, 
and on the shape function~\cite{berne}
\begin{eqnarray}
\lefteqn{\sigma_{12}(\uu_1,\uu_2,\ru_{12}) =
\sigma_0 \: \left\{ 1 - \frac{\chi}{2} \left[
\frac{(\uu_1\cdot\ru_{12}+\uu_2\cdot\ru_{12})^2}
     {1+\chi \uu_1\cdot\uu_2} \right. \right.
}\qquad \qquad \nonumber \\&& \left. \left. + 
\frac{(\uu_1\cdot\ru_{12}-\uu_2\cdot\ru_{12})^2}
     {1-\chi \uu_1\cdot \uu_2} \right] \right\}^{-1/2},
\end{eqnarray}
which approximates the contact distance between two ellipsoids of
elongation $\kappa =  \sqrt{(1+\chi)/(1-\chi)}$ with orientations 
$\uu_1$ and $\uu_2$ and center-center vector $\rr_{12}$ pointing in
the direction $\ru_{12}=\rr_{12}/r_{12}$. The parameters were 
$\kappa = 3$ and $k_B T = 0.5 \epsilon_0$. 

Four state points were considered, two at densities $\varrho = 0.24/\sigma_0^3$
and $0.283/\sigma_0^3$ in the isotropic phase, and two at densities
$0.287/\sigma_0^3$ and $\varrho = 0.3/\sigma_0^3$ in the nematic phase. 
The system size was $N=1000$ particles at $\varrho = 0.24/\sigma_0^3$,
$N=4000$ at $\varrho = 0.283/\sigma_0^3$ and $0.287/\sigma_0^3$, and
$N=1000-8000$ at $\varrho=0.3/\sigma_0^3$. The systems with the three
lower densities were studied in the canonical ensemble by Monte Carlo 
simulation. The data for the larger systems ($N \ge 4000$) at the density
$\rho = 0.3 \sigma^{-1}$ were originally produced by 
G.~Germano~\cite{phuong1}. They were simulated in the microcanonical ensemble 
using a parallel molecular dynamics program on a CRAY T3E. Here the moment 
of inertia of a particle was chosen $I=2.5 m_0 \sigma_0{}^2$ and the time 
step $\Delta t = 0.003 \sigma_0 \sqrt{m_0/\epsilon_0}$,
where $m_0$ is the mass of one particle. The results did not depend
on the simulation method. Run length were 5-10 million MC or MD steps, 
depending on the system size, and data were collected every 
1000 or 10000 steps.

Throughout this paper, the units will be defined in terms of
the energy unit $\epsilon_0$, the mass unit $m_0$, the
temperature unit $\epsilon_0/k_B$, and the length unit $\sigma_0$.

The orientational order is characterized as usual by the 
order tensor~\cite{degennes}
\begin{equation}
\label{eq:op}
\QQ = \langle \frac{1}{N} \sum_{i=1}^{N}
(\frac{3}{2} \uu_i \otimes \uu_i - \frac{1}{2} {\II} )
\rangle,
\end{equation}
where the sum $i$ runs over all $N$ particles of orientation $\uu_i$,
$\II$ is the unit matrix, $\otimes$ the dyadic vector product,
$\langle \cdot \rangle$ denotes thermal averages. The largest eigenvalue 
of $\QQ$ is the nematic order parameter $V P_2$, and the corresponding 
eigenvector $\nn$ is the director, which points along the direction 
of prefered alignment. In our case, we have $P_2 = 0.47$ at 
$\varrho=0.287/\sigma_0^3$, and $P_2 = 0.69$ at $\varrho=0.3/\sigma_0^3$. 
The densities $\varrho=0.283/\sigma_0^3$ and $\varrho=0.287/\sigma_0^3$
are close to the coexisting densities at the isotropic/nematic phase
transition. Previous constant pressure simulations of the same 
system~\cite{harald} showed that the transition is only very weakly
first order, and the width of the coexistence gap is of the order
of $0.05 /\sigma_0^3$.

\subsection{Data analysis}
\label{sec:data}

Before describing our way to analyze the data, we recall some basic 
definitions: In a system of particles $i$ with orientation $\uu_i$ 
and position $\rr_i$, the one-particle distribution can be written 
as~\cite{hansen}
\begin{equation}
\rho^{(1)}(\uu,\rr) = \langle \sum_i 
\delta(\uu - \uu_i) \delta(\rr - \rr_i) 
\rangle.
\end{equation}
and the pair distribution as
\begin{eqnarray}
\lefteqn{\rho^{(2)}(\uu_1,\rr_1,\uu_2,\rr_2)} \\
&& = \langle \sum_{i\ne j}
\delta(\uu_1 - \uu_i) \delta(\rr_1 - \rr_i) 
\delta(\uu_2 - \uu_j) \delta(\rr_2 - \rr_j) 
\rangle.\nonumber
\end{eqnarray}
In homogeneous nematic fluids, the pair distribution depends on the
difference vector $\rr_{12} = \rr_1-\rr_2$ rather than on the 
individual positions $\rr_1$ and $\rr_2$, and the 
single-particle distribution depends only
on $\uu$ (in fact, on $|\uu \cdot \nn|$). In that case, the total 
correlation function $h(\uu_1,\uu_2,\rr_{12})$ is defined by
\begin{equation}
\label{eq:TCF}
h(\uu_1,\uu_2,\rr_{12}) = \frac{\rho^{(2)}(\uu_1,\uu_2,\rr_{12})}
{\rho^{(1)}(\uu_{1}) \rho^{(1)}(\uu_{2})} -1,
\end{equation}
and the direct correlation function $c(\uu_1,\uu_2,\rr_{12})$ is determined
by the Ornstein-Zernike equation~\cite{hansen,gray,workman}
\begin{eqnarray}
\label{eq:OZ}
\lefteqn{h({\uu_1,\uu_2,\rr_{12}}) = c(\uu_1,\uu_2,\rr_{12}) +}
& \qquad &
\nonumber\\ &&
\int c(\uu_1,\uu_3,\rr_{13}) \: \rho^{(1)}(\uu_3) \:
 h(\uu_3,\uu_2,\rr_{32}) d\uu_{3} d\rr_{3}.
\end{eqnarray}

We choose a coordinate frame where the $z$ axis points in the direction
of the director $\nn$ (director frame)~\cite{footnote1}. 
All orientation dependent functions are 
expanded in spherical harmonics $Y_{lm}(\uu)$. This yields
\begin{equation}
\label{eq:fl}
\rho^{(1)}(\uu) = \varrho \sum_{l \: \mbox{\tiny even} } f_{l} \:Y_{l
0}(\uu),
\end{equation}
with the bulk number density $\varrho$, and
\begin{eqnarray}
F(\uu_1,\uu_2,\rr)& =& \! \! \!
\sum_{{{l_1,l_2,l}\atop {m_1,m_2,m}}} \!\!
F_{l_1 m_1 l_2 m_2 l m}(r) \:
\label{eq:Fl}
\\
&& \qquad \qquad
Y_{l_{1}m_{1}}(\uu_1) \: Y_{l_{2}m_{2}} (\uu_2) \: Y_{lm}(\ru),
\nonumber
\end{eqnarray}
where $F$ stands for $\rho^{(2)}$, $h$ or $c$, and $\ru$ denotes
the unit vector $\rr/r$. In uniaxially symmetric phases, 
only real coefficients with $m_1 + m_2 + m = 0$ and even $l_1 + l_2 + l$
enter the expansion. Since our particles have uniaxial symmetry,
every single $l_i$ is even as well.

The expansion coefficients of the pair distribution function
can be determined directly from the simulations~\cite{streett}
\begin{eqnarray}
\lefteqn{\rho^{(2)}_{l_1 m_1 l_2 m_2 l m}(r) = 4 \pi \: \varrho^2 \:
g(r)}
\qquad \qquad \nonumber\\
\label{eq:streett}
&&
\langle
\: Y^*_{l_1 m_1}(\uu_1) Y^*_{l_2 m_2}(\uu_2) Y^*_{l m}(\ru) \:
\rangle_{\delta r},
\end{eqnarray}
where the average $\langle \cdot \rangle_{\delta r}$ is performed
over all molecule pairs at distances $|\rr_1-\rr_2| \in [r,r+\delta r]$,
and $g(r)$ is the radial distribution function, i. e., the average
total number of molecule pairs divided by $ N 4 \pi \varrho r^2 \delta r$.
The coordinate frame was determined separately in each configuration,
such that the $z$ axis is given by the direction of the director
(director frame). In general, we have determined coefficients for values of 
$l,l_i$ up to $l_{\mbox{\tiny max}}=6$. For the highest density~\cite{phuong1},
we checked in small systems that the truncation is sufficient by comparing 
the results with those obtained using $l_{\mbox{\tiny max}}=8$.
The bin size was $\delta r = 0.04 \sigma_0$ and only ranges up to
$r_{\mbox{\tiny max}} = 0.4 L$ ($L$ being the size of the simulation
box) were considered in order to reduce boundary effects~\cite{pratt}.

The total correlation function $h$ is obtained from the
spherical harmonics version of eqn. (\ref{eq:TCF}),
which is simply a matrix equation:
\begin{eqnarray}
\lefteqn{
\rho^{(2)}_{l_1 m_1 l_2 m_2 l m}(r) =
\varrho^{2} \Big(\:
\sqrt{4 \pi} f_{l_1} f_{l_2}
\delta_{m_1 0}\delta_{m_2 0}\delta_{l 0} \delta_{m 0}
}\quad 
 \\&&
\label{eq:TCF_2}
+ \:
\sum_{{l_1' l_1'',l_2',l_2''}}
h_{l_1' m_1 l_2' m_2 l m}(r)
f_{l_1''} f_{l_2''}
\textstyle
\Gamma^{l_1 \, l_1' \, l_1''}_{m_1 m_1 0}
\Gamma^{l_2 \, l_2' \, l_2''}_{m_2 m_2 0} \:
\Big), \nonumber
\end{eqnarray}
with
\begin{equation}
\Gamma^{l \: l' l''}_{m m' m''} =
\int d\uu \: Y_{lm}^*(\uu) Y_{l',m'}(\uu) Y_{l'',m''}(\uu) 
\end{equation}
The DCF $c$ is most easily calculated in Fourier representation.
The expansion coefficients of a function $F(\uu_1,\uu_2,\rr)$ in
real space are related to their counterparts in Fourier space
by the Hankel transform~\cite{gray}
\begin{eqnarray}
\label{eq:hankel}
F_{l_1 m_1 l_2 m_2 l m}(k) &=& 4\pi i^l 
\!\! \int_{0}^{\infty}\!\!\!\!\!
 r^2\, j_l(kr) \, F_{l_1 m_1 l_2 m_2 l m}(r)\: dr \\
F_{l_1 m_1 l_2 m_2 l m}(r) &=& \frac{4\pi (-i)^l }{(2 \pi)^3}
\!\! \int_{0}^{\infty}\!\!\!\!\!
 k^2\, j_l(kr) \, F_{l_1 m_1 l_2 m_2 l m}(k)\: dk 
\end{eqnarray}
where $j_l(kr)$ is the spherical Bessel function. 
The Ornstein-Zernike equation (\ref{eq:OZ}) in Fourier space 
and spherical harmonics representation reads
\begin{eqnarray}
h_{l_1 m_1 l_2 m_2 l m }(k) & = &
c_{l_1 m_1 l_2 m_2 l m}(k) \nonumber \\
\lefteqn{
+ \varrho \sum_{{l_3 l_3' l_3'' m_3}\atop {l' m' l'' m'' l_3}}\!\!
c_{l_1 m_1 l_3 m_3 l' m'}(k) \:
h_{l_3' m_3 l_2 m_2 l'' m''}(k) \:
} \qquad \qquad
\nonumber\\ &&
\label{eq:OZ_2}
\times f_{l_3''} 
(-1)^{m_3}
\Gamma^{l l' l''}_{m m' m''}
\Gamma^{l_3 l_3' l_3''}_{m_3 m_3 0},
\end{eqnarray}
which is again a matrix equation.
\end{multicols}\twocolumn

Special care must be taken when performing the Hankel transform 
(\ref{eq:hankel}) in the nematic state: Due to the elasticity of
the nematic fluid, certain coefficients of $h$ (in particular those with 
$m_1, m_2 = \pm 1$) have pronounced long-range tails and exhibit
a $1/r$ behavior at large distances $r$. Furthermore, large-wavelength
fluctuations are suppressed in finite systems, and the director
frame follows the local order if evaluated in a small box.
As a result, some coefficients (those with $l=0$, $m_i=\pm1$) 
are shifted towards zero in finite systems by an amount which 
is almost constant beyond $r = \sigma_0$ (see Figure \ref{fig:shift}). 
Before performing the Hankel transform, we thus fit the data points 
beyond distances $r>r_0$ to a power law $a + b/r$, shift $h(r)$ by $a$, 
and extrapolate it to infinity. At the highest density 
$\varrho = 0.3/\sigma_0^3$, we have simulated systems of different size 
($N=1000, N=4000$, and $N=8000$) and checked that finite
size effects are basically eliminated by our procedure. 
The parameter $r_0$ was chosen $r_0=4.0/\sigma_0$ at $N=4000$,
$r_0=2.8/\sigma_0$ at $N=1000$, and $r_0=5.3/\sigma_0$ at $N=8000$.

\fig{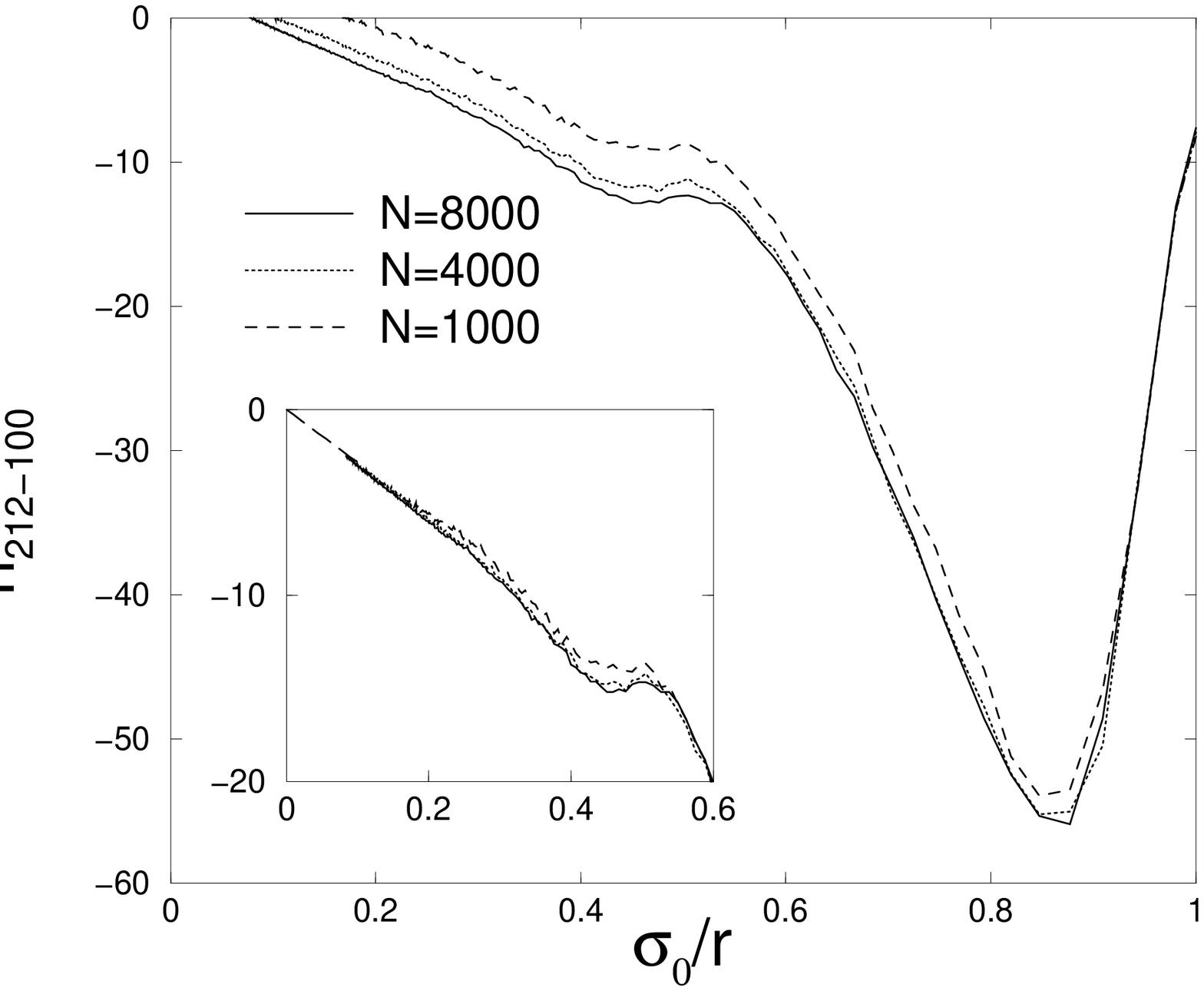}{70}{70}{}
\noindent
\begin{figure}[t]
\CCshift
\end{figure}

The data manipulations which finally yield the DCF are altogether quite
extensive. An independent test of the results which checks the procedure 
is clearly desirable. For example, one can calculate the elastic constants 
from the DCF using the Poniewierski-Stecki relations~\cite{ps}, and 
compare them with values obtained independently from the structure factor. 
As we have reported in our earlier paper~\cite{phuong1}, the agreement
is quite satisfactory. As another test, we have verified the validity
of a relation originally derived by Gubbins~\cite{gubbins},
\begin{equation}
\label{eq:gubbins}
\frac{\rho^{(1)}{}'(u_{1,z})}
{\rho^{(1)}(u_{1,z})} \: u_{1,\alpha} 
=
\int  c(\uu_1,\uu_2,\kk)|_{k=0} \:
\rho^{(1)}{}'(u_{2,z}) u_{2,\alpha} \: d\uu_2.
\end{equation}
In Fourier space and spherical harmonics representation, this reads
\begin{eqnarray}
\label{eq:gubbins2}
\lefteqn{
\sqrt{l(l+1)} \int \ln(\rho^{(1)}(\uu)) \: Y_{l0}(\uu) \: d\uu
} \\
&=& 
-\frac{\varrho}{2\sqrt{\pi}} \sum_{l'} \!\!
\sqrt{l'(l'+1)} f_{l'} c_{l 1 l' -1 0 0}(k)|_{k=0}.
\nonumber
\end{eqnarray}
The quantity on the right hand side of this equation is plotted
as a function of $k^2$ for $l=2$ and $l=4$ in Figure \ref{fig:gubbins}. 
The $k \to 0$ limit agrees well with the value obtained from 
the left hand side.

\fig{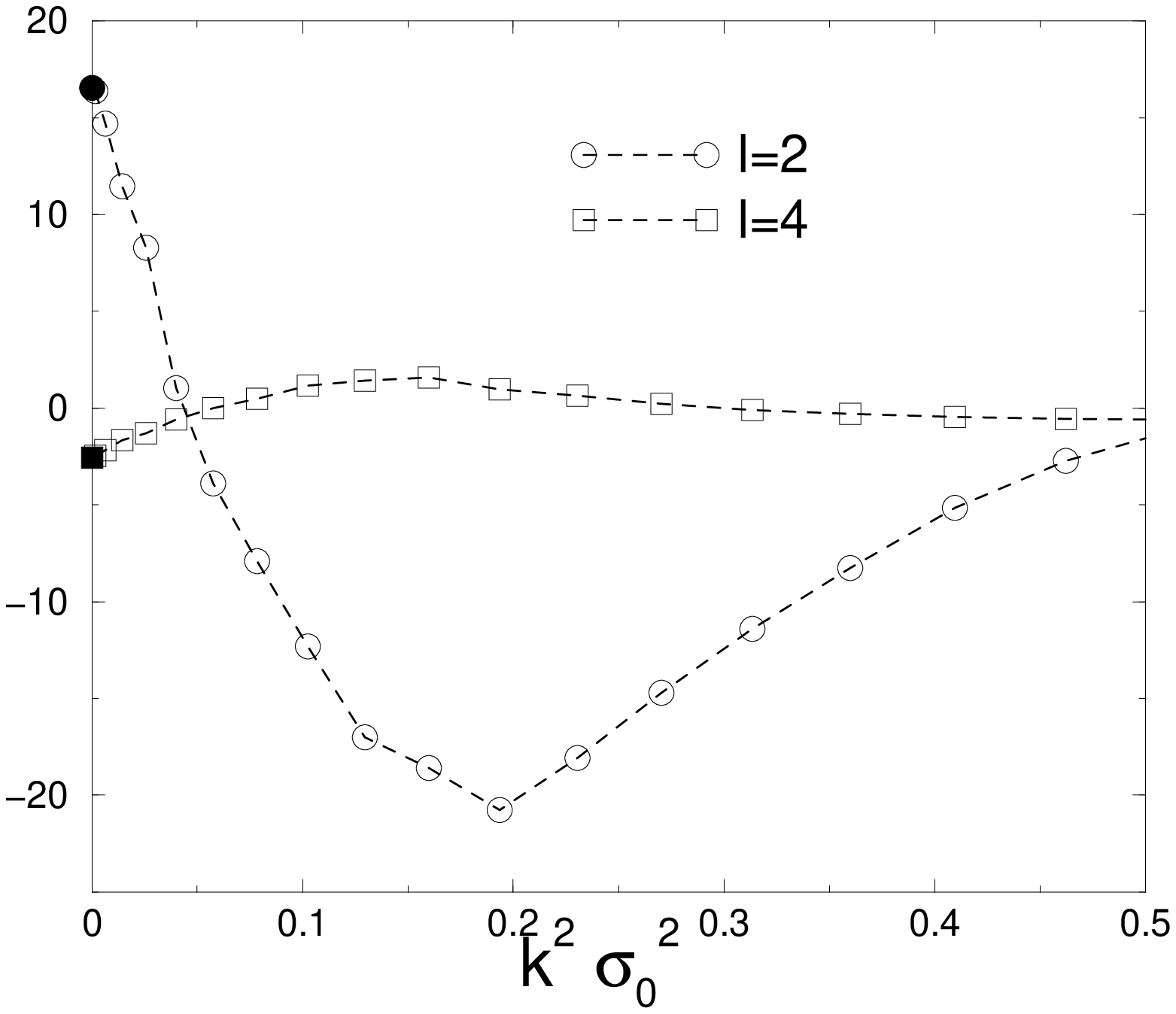}{70}{70}{}
\begin{figure}[t]
\CCgubbins
\end{figure}

\section{Results}
\label{sec:results}

The orientational average of the pair correlation functions is 
shown in Figure \ref{fig:aver} for different densities. It is
basically given by their respective spherical harmonics coefficients 
with $l,l_i,m,m_i = 0$. The pair distribution function $\rho^{(2)}$ 
exhibits several peaks at distances slightly larger than $\sigma_0$,
corresponding to the nearest neighbor shell, the next
nearest neighbor shell etc. The peaks are less pronounced,
but still present in the total correlation function.
They disappear in the DCF. At low density, the DCF vanishes
at distances beyond $\kappa \sigma_0$. At large densities,
however, one observes broad oscillations which persist further. 
Orientationally averaged, the structure of the DCF does not appear 
to be much simpler than that of the pair distribution function. 

~

\fig{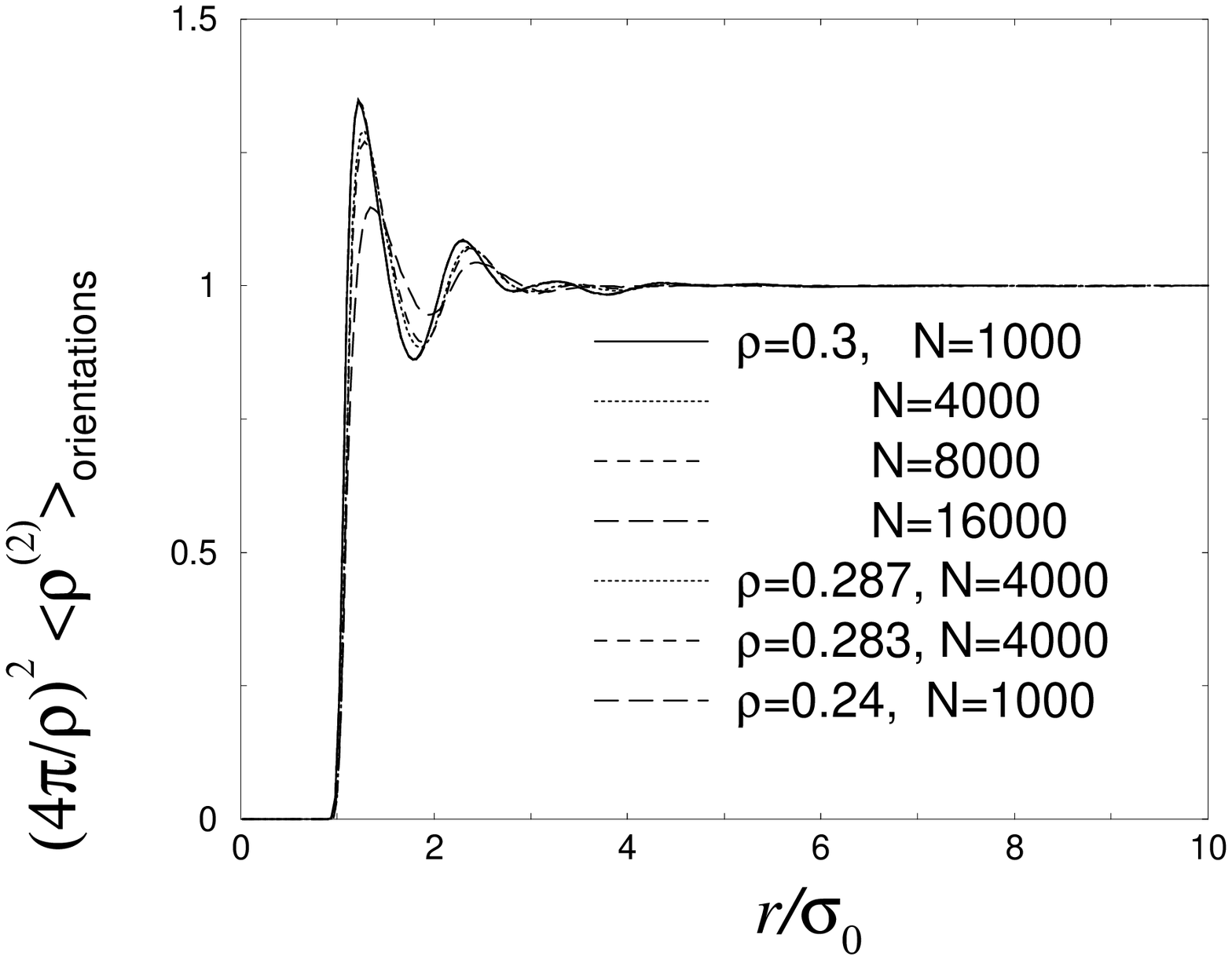}{70}{57}{(a)}

\fig{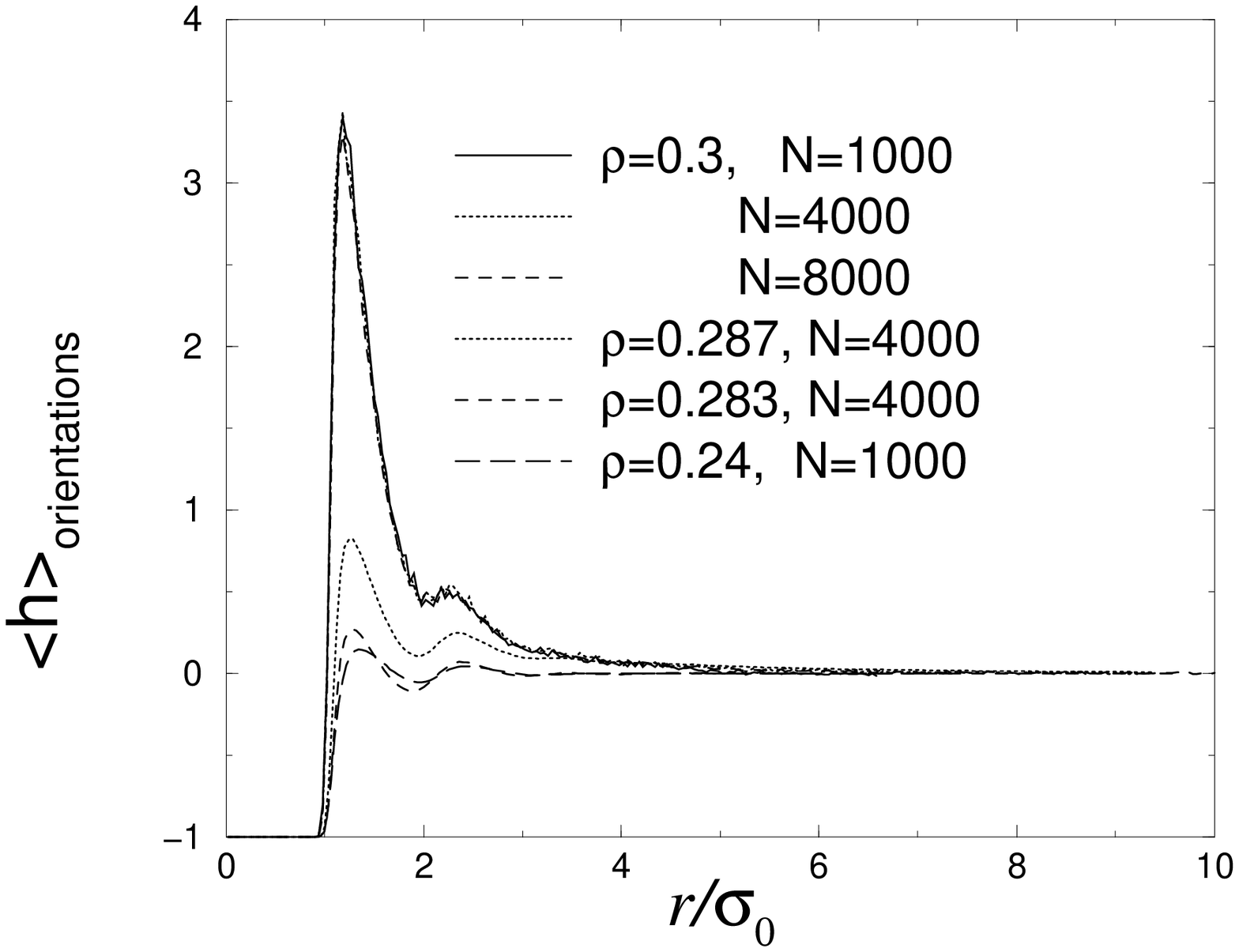}{70}{55}{(b)}

\fig{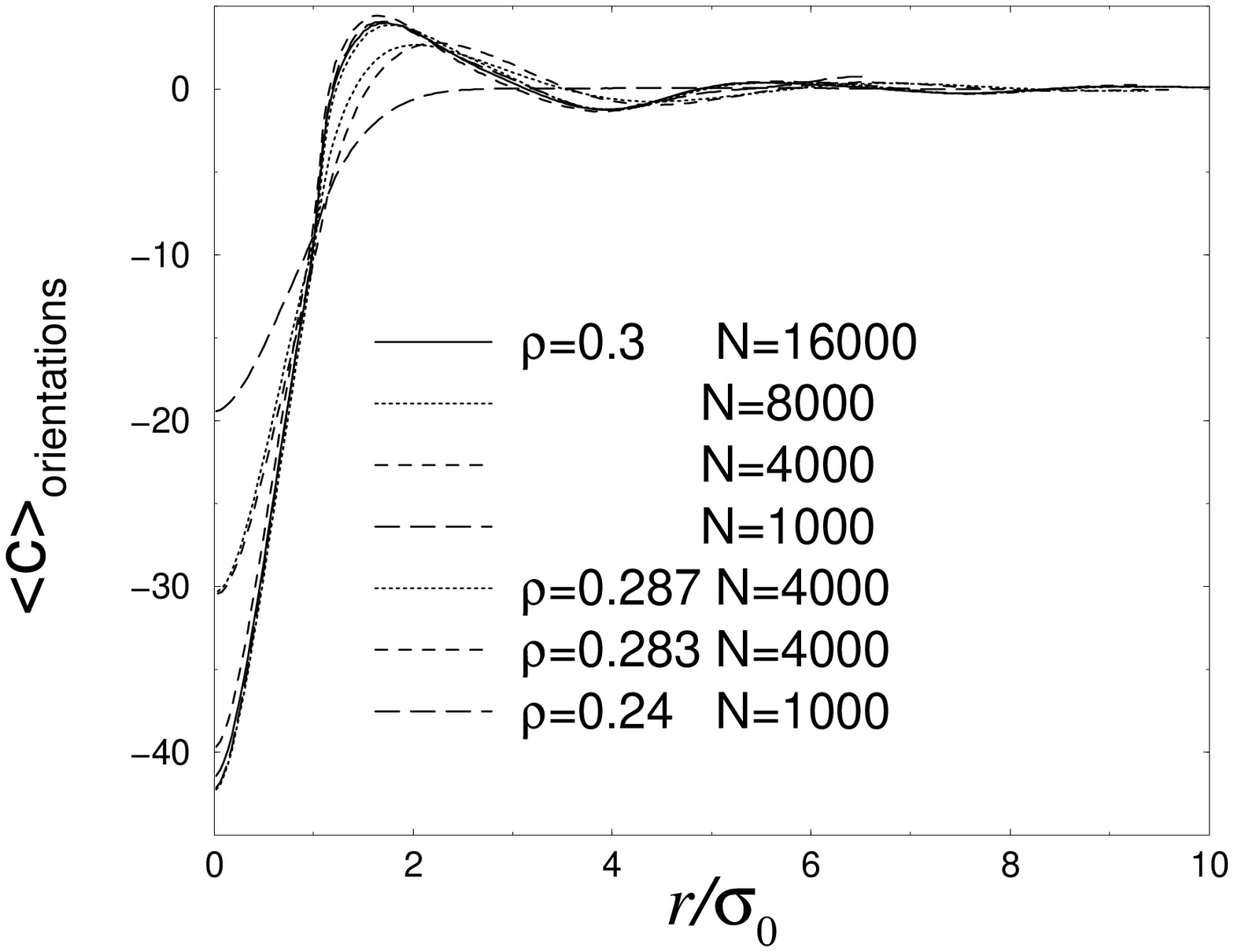}{70}{55}{(c)}

\begin{figure}[t]
\CCaver
\end{figure}

The qualitative difference between the DCF and the total correlation
function becomes apparent when looking at coefficients which reflect 
the elasticity of the nematic fluid. Figure \ref{fig:elastic} shows 
as an example the coefficients with $l_1 = l_2 = l = 2$, $m_1 = -m_2 = 1$ 
and $m = 0$. In the nematic phase, they decay at long distance like $1/r$, 
both for the pair distribution function and the total pair cor-

~

\fig{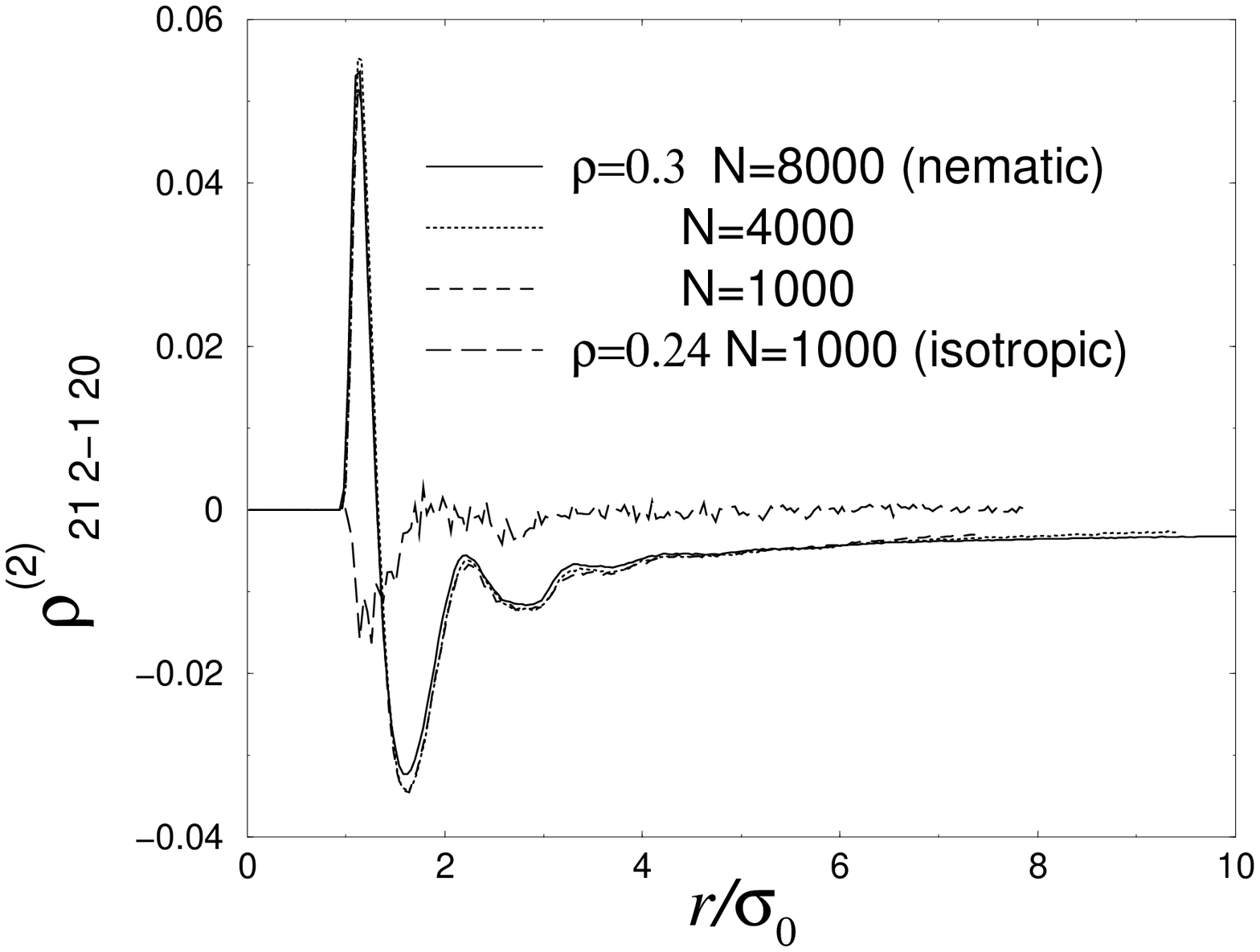}{70}{57}{(a)}

\fig{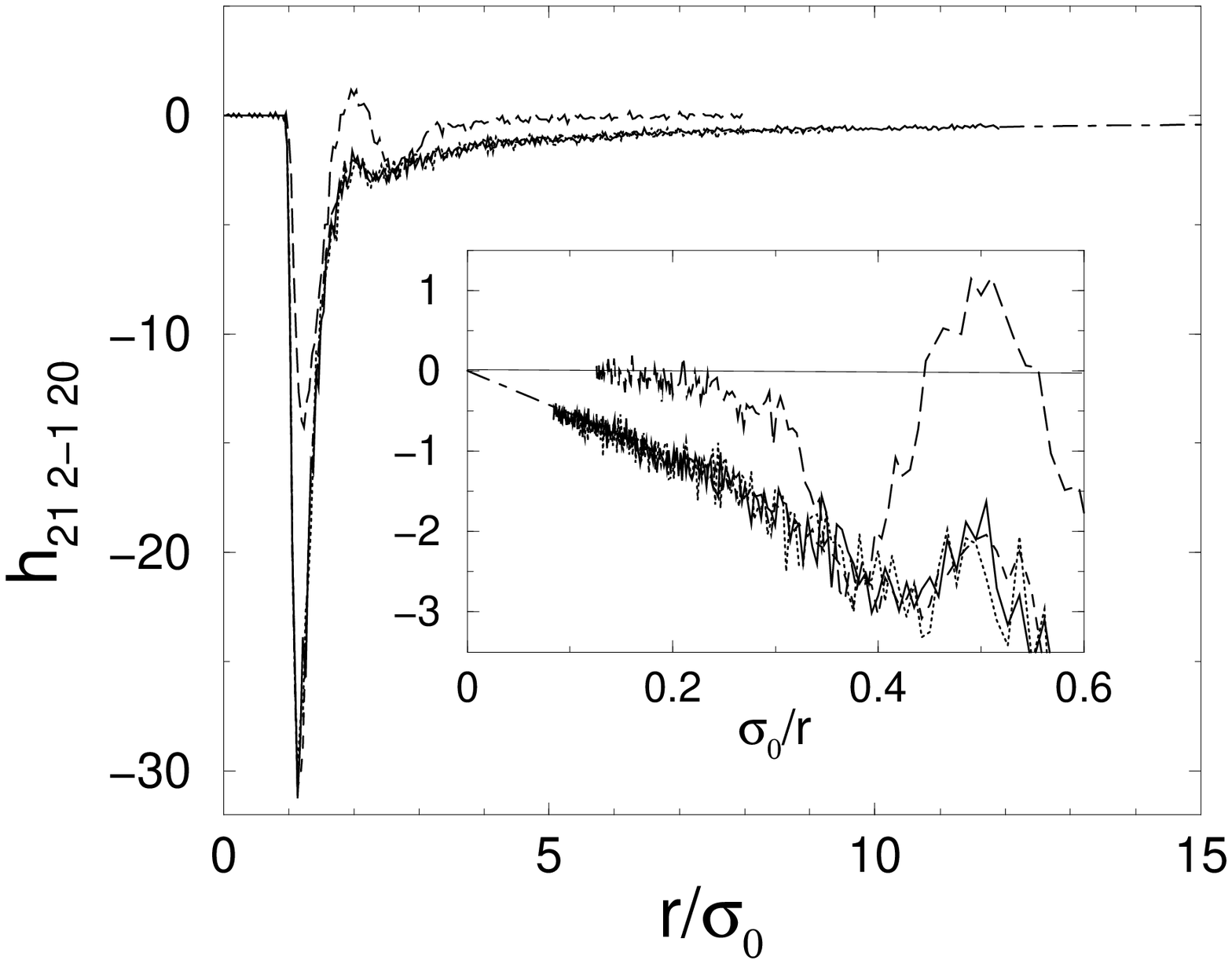}{70}{55}{(b)}

\fig{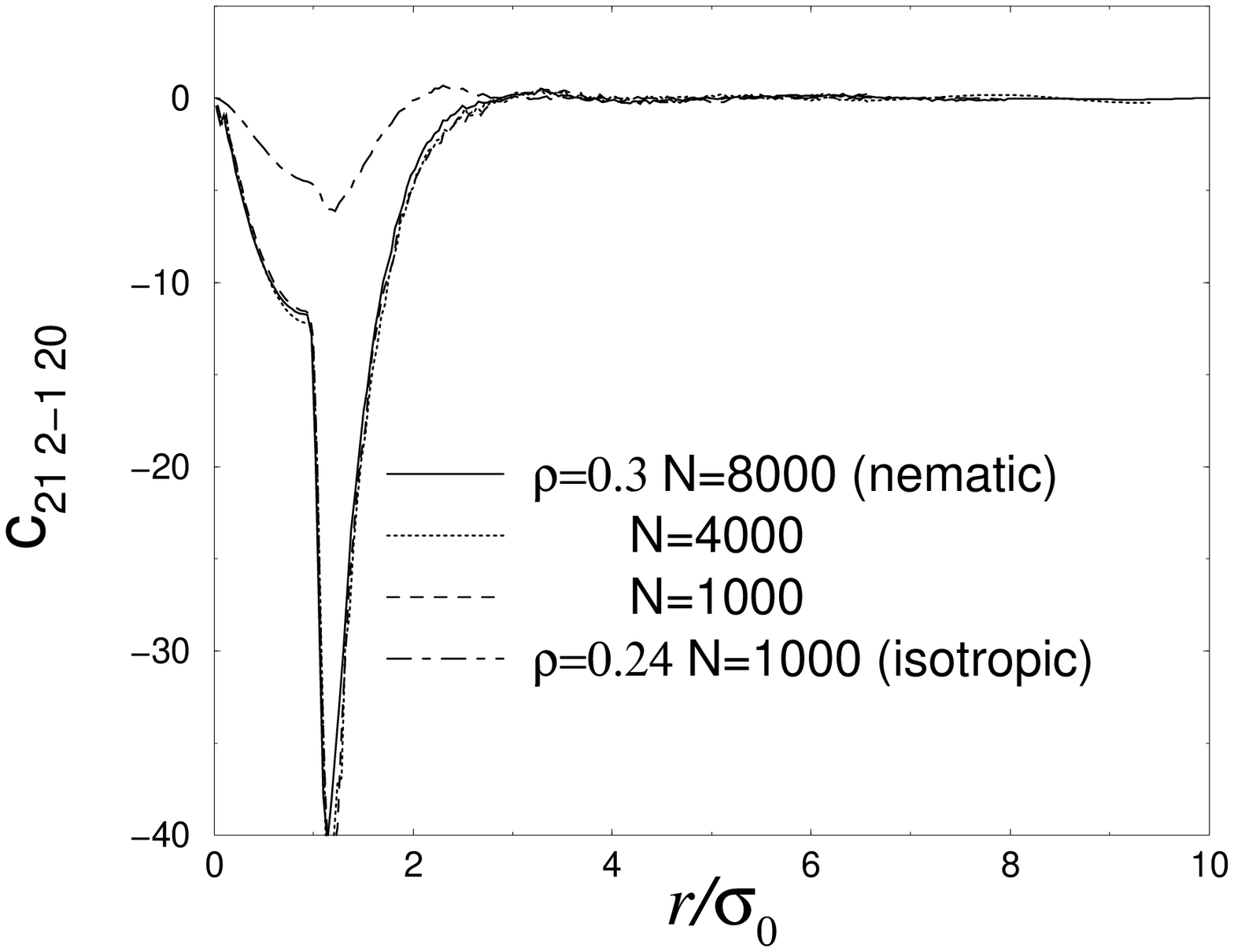}{70}{55}{(c)}

\begin{figure}[t]
\CCelastic
\end{figure}

\noindent
relation function. The DCF no longer exhibits such a long ranged tail.
Thus the DCF is indeed a quantity which characterizes the structure
of nematic fluids on a local scale.

 ~

\fig{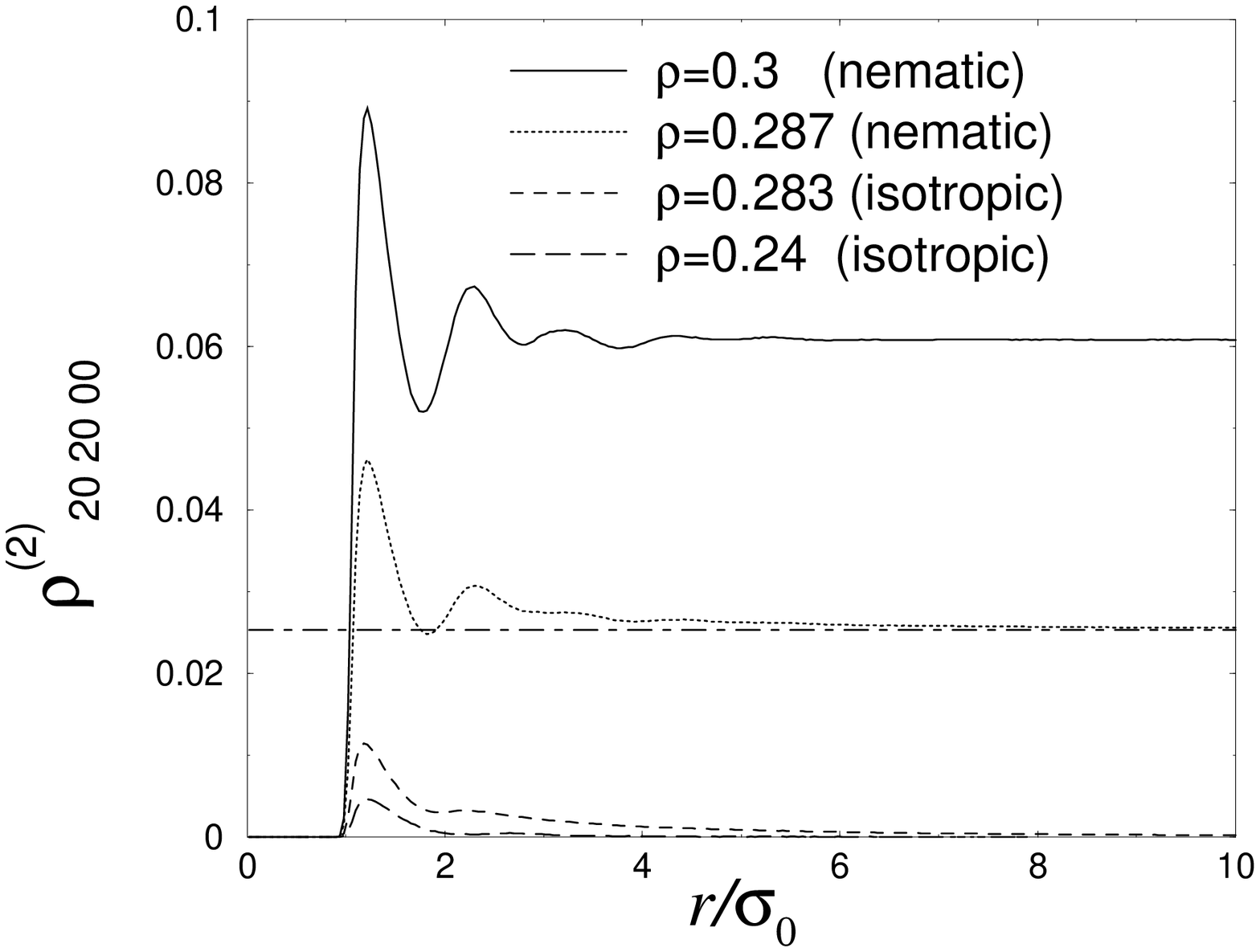}{70}{57}{(a)}

\fig{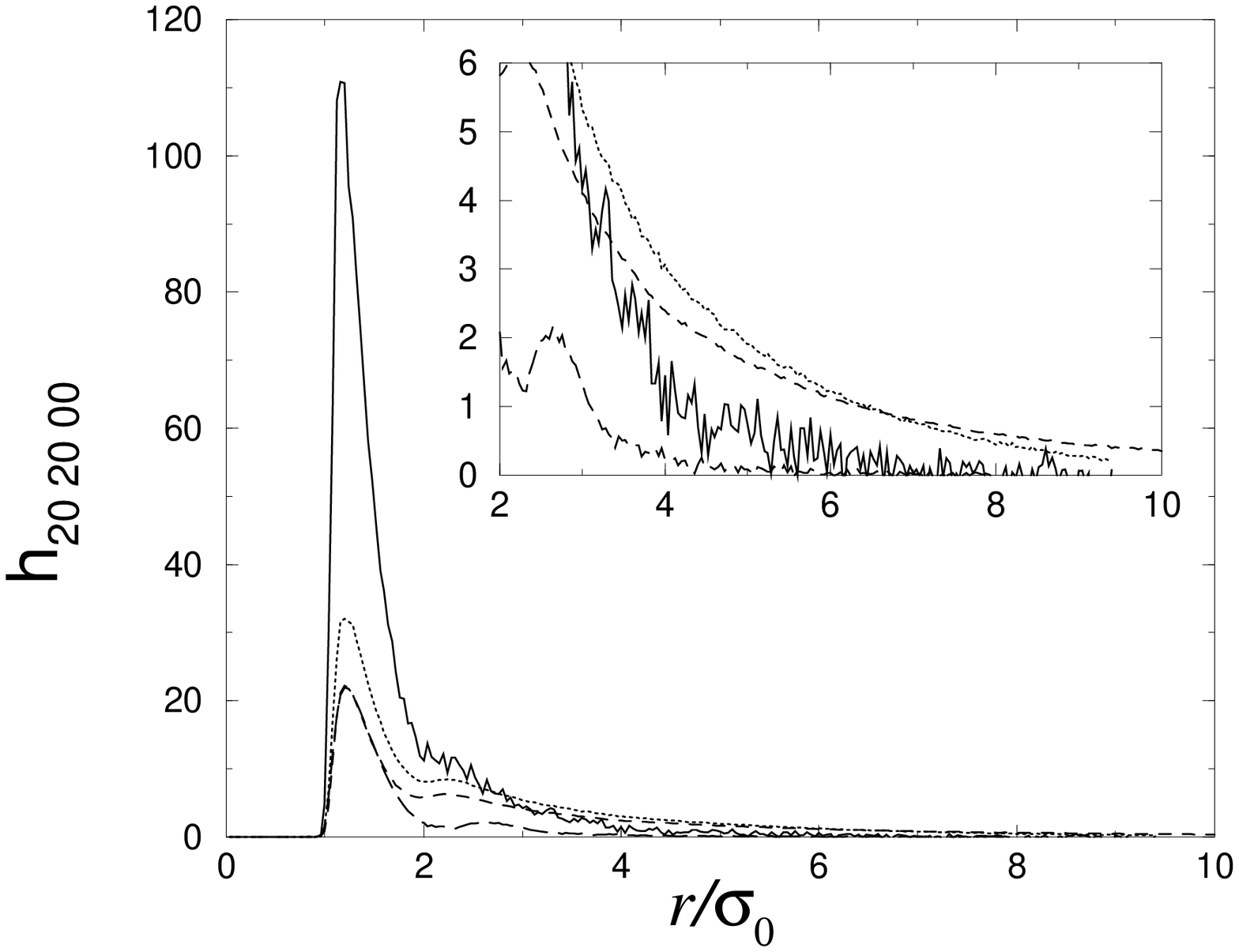}{70}{55}{(b)}

\fig{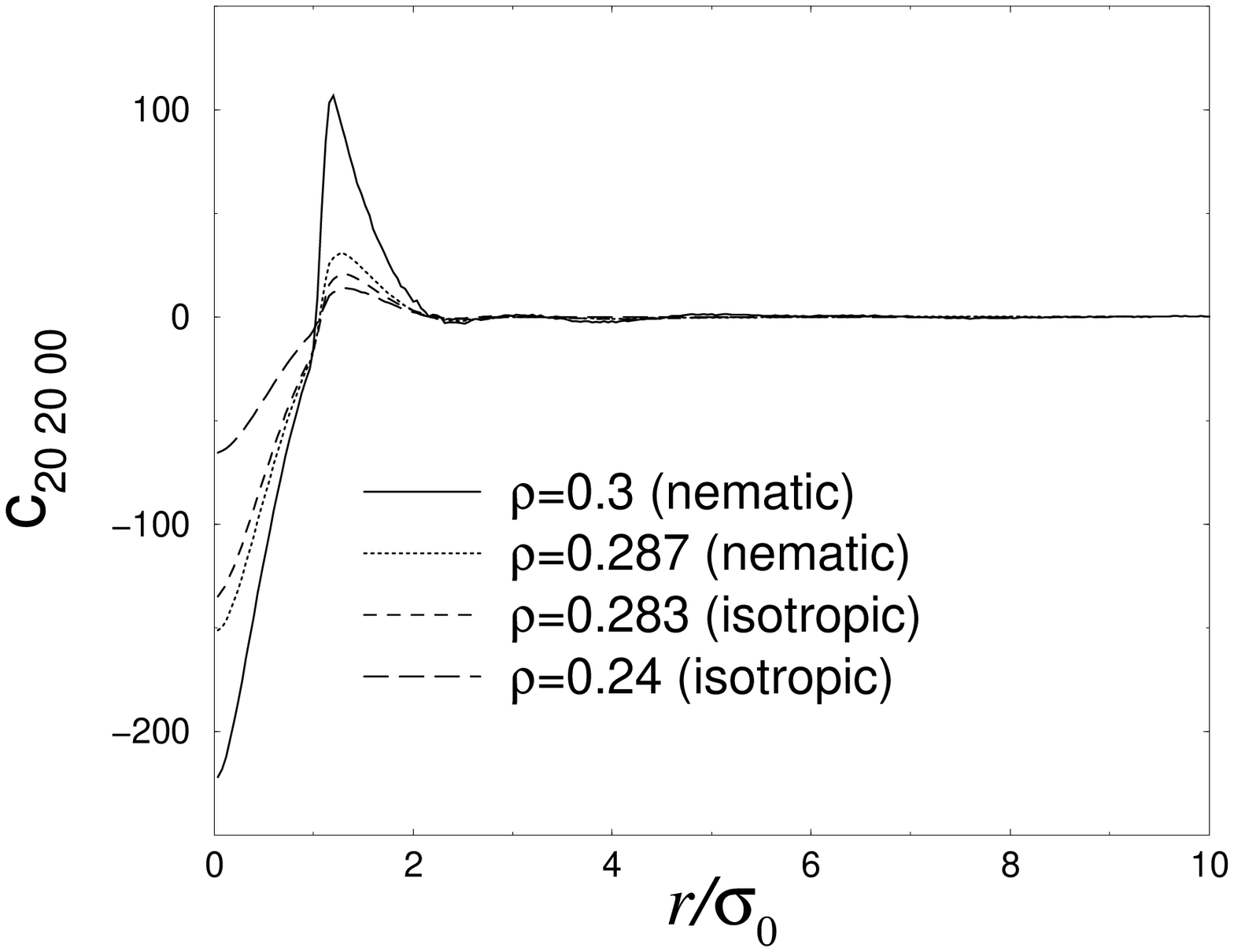}{70}{55}{(c)}

\begin{figure}[t]
\CCorient
\end{figure}

Next we investigate the isotropic/nematic transition in more detail. 
Figure \ref{fig:orient} illustrates how orientational order emerges and
gives rise to a constant long range contribution to the pair distribution 
function coefficients with $l_i \ne 0, m_i=l=m=0$. The constant contribution 
is subtracted in the total correlation function (cf. eqn. (\ref{eq:TCF})). 
Close to the transition, however, the total correlation function features 
a slowly decaying, quasi long range tail (Fig. \ref{fig:orient} b). 
It stems from near critical fluctuations with large correlation lengths 
close to the transition, due to the fact that the transition is only weakly
first order. Like the elastic tails, these quasicritical long range tails
are no longer present in the DCF.

From the results shown so far (Figs \ref{fig:aver}, \ref{fig:elastic}, 
and \ref{fig:orient}), the effect of nematic ordering on the DCF is not 
obvious. The influence of nematic symmetry breaking on correlation functions 
is best studied in a molecular-fixed frame system, which takes full 
advantage of the symmetries in the isotropic phase and thus allows to 
assess directly the loss of symmetries in the nematic phase. Following 
Gray and Gubbins~\cite{gray}, we introduce a local coordinate system 
in which the $z$ axis points in the direction ${\bf r}_{12}$ of the 
intermolecular distance vector (``molecular frame'', see figure 
\ref{fig:frames}).
The other two axes are chosen such that the $x$ axis lies in the plane 
spanned by $\rr_{ij}$ and the director $\nn$~\cite{footnote1}.

\hspace*{1cm} \fig{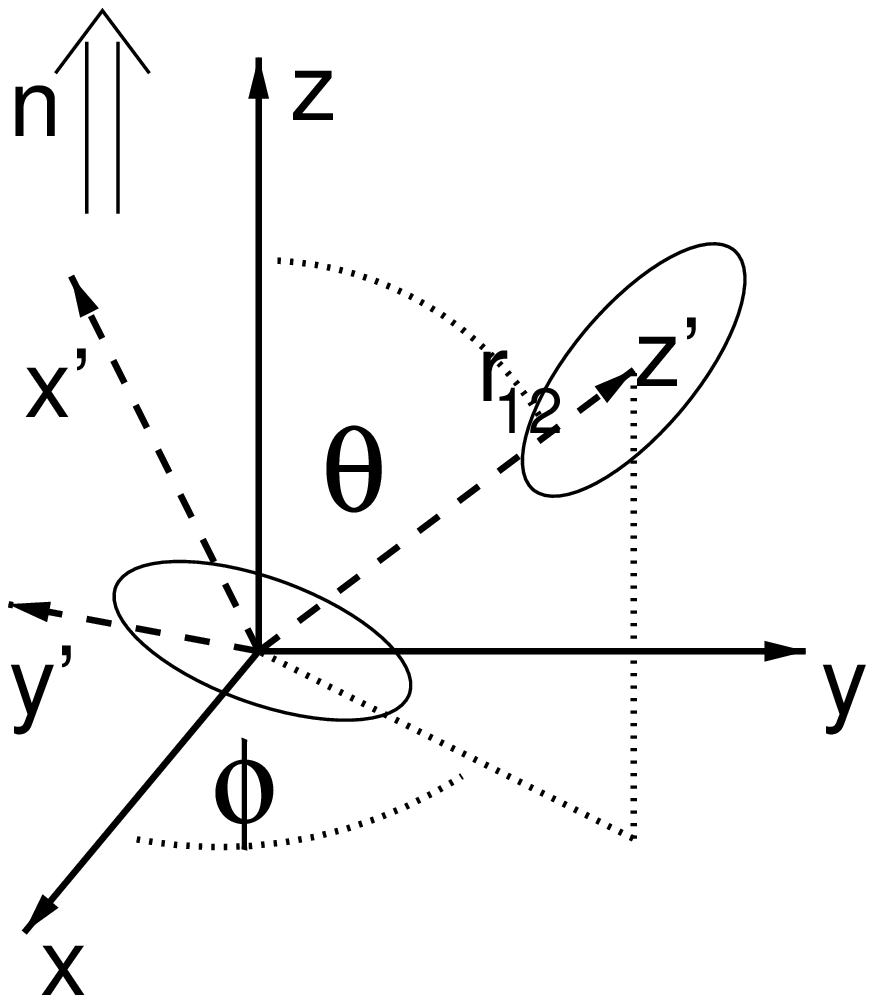}{48}{58}{}
\begin{figure}
\CCframes
\end{figure}

This frame is now used in the spherical harmonics expansion (\ref{eq:Fl}) of 
functions which depend on particle orientations $\uu_i$. The intermolecular 
distance vector $\rr_{12}$ itself is still represented in the director frame.
In an isotropic fluid, pair correlation functions must be rotationally
invariant, and all expansion coefficients except those with $l=m=0$ vanish.
In a nematic fluid, this is no longer true~\cite{footnote2}. 
Rotational invariance is broken if coefficients with $l \ne 0$ become nonzero. 
One would certainly expect this to happen for the pair distribution function 
and the total pair correlation function. In the case of the DCF, the situation 
is less clear. Zhong and Petschek \cite{zhong1,zhong2} have argued that 
the DCF should remain rotationally invariant in an aligned fluid,
since it reflects effective pair potentials between particles. 

\fig{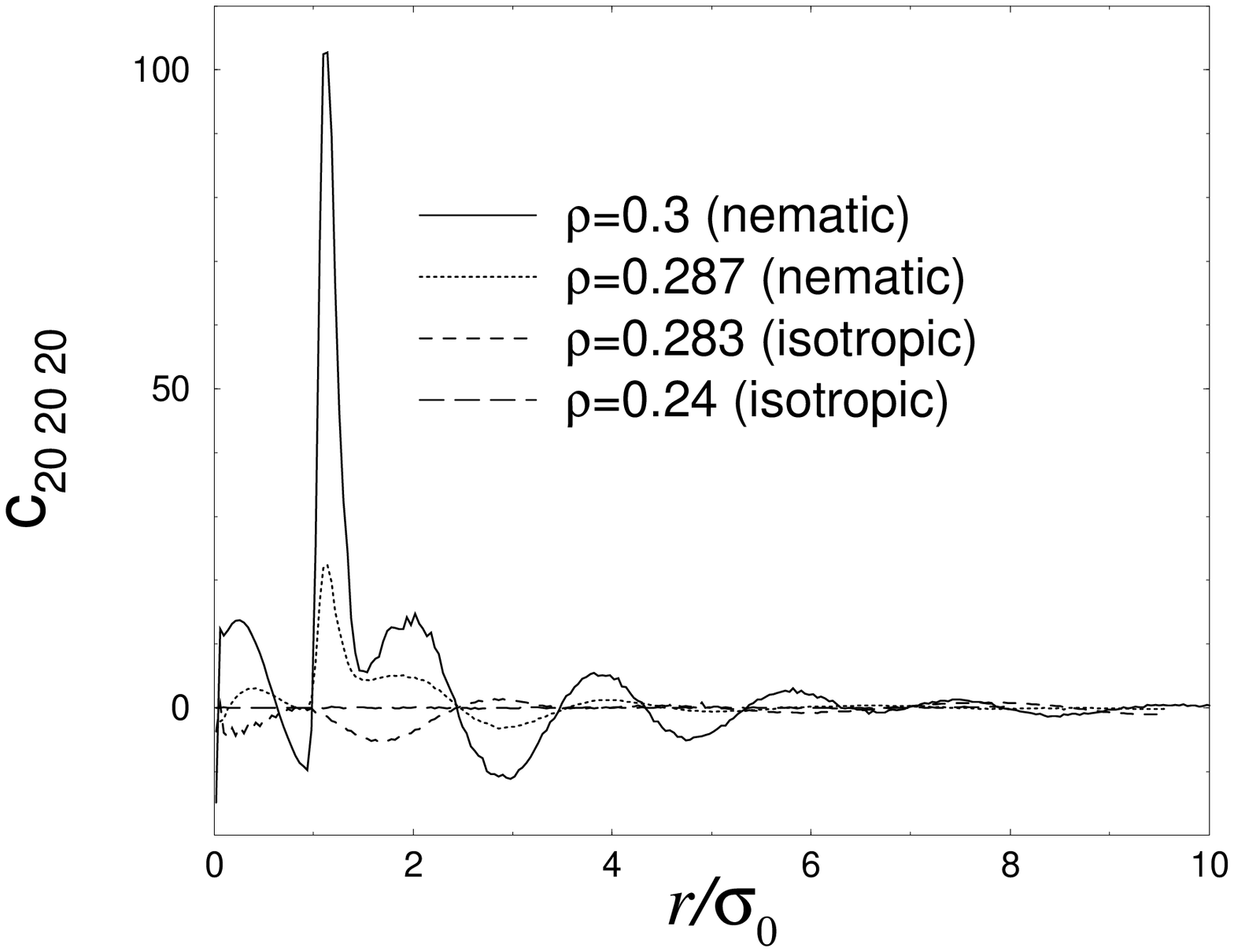}{70}{65}{}
\begin{figure}
\CCnibreak
\end{figure}

\fig{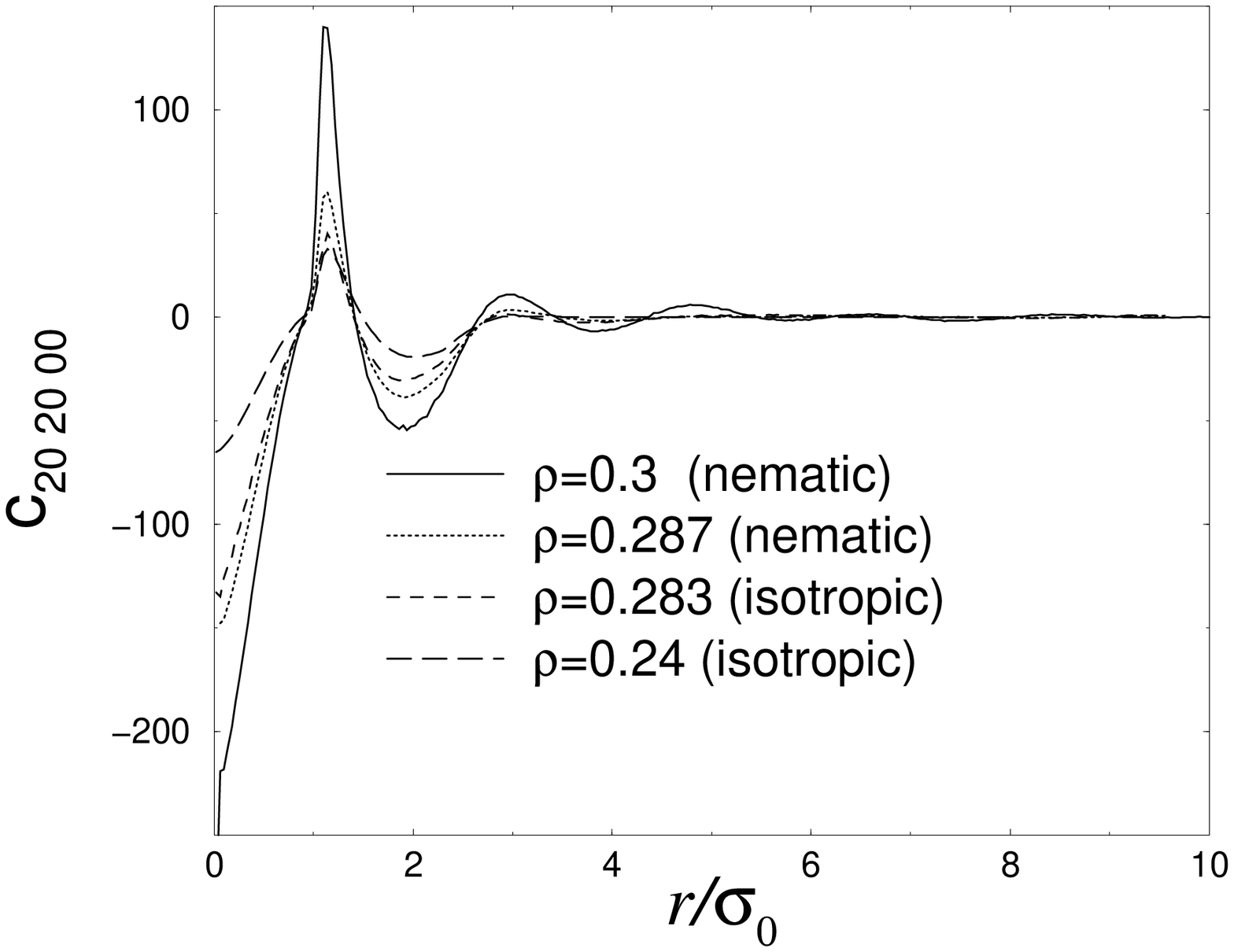}{70}{65}{}
\begin{figure}
\CCnisym
\end{figure}

An equation which converts expansion coefficients from the director 
frame into the molecular frame is derived in the appendix. 
Such conversion formulae can be found for isotropic fluids in 
the literature~\cite{gray}. Our general case is more involved.

Figure \ref{fig:nibreak} shows an example of an expansion coefficient,
$c_{202020}$, which vanishes in the case of rotational invariance. 
In the isotropic phase, the values are indeed close to zero. In the
nematic phase, a nonzero DCF coefficient not only emerges, it
may even grow quite large. At $\rho = 0.3 \sigma_0^{-3}$ the values
are comparable to those of a symmetry preserving coefficient of
similar order (\ref{fig:nisym}). Hence the DCF reflects
the broken symmetry in the nematic phase. Zhong's and Petschek's 
seemingly reasonable assumption is not valid for fluids of ellipsoids.

In contrast, the symmetry preserving coefficient $c_{202000}$ seems
barely affected by the transition (Fig. \ref{fig:nisym}). The curves
for $\rho=0.283 \sigma_0^{-3}$ and $\rho=0.287 \sigma_0^{-3}$,
slightly below and above the transition, lie almost on top of each 
other.

It is instructive to compare the simulation results for the DCF
with the results from the Percus-Yevick theory~\cite{percus}. 
In the Percus-Yevick approximation, the direct correlation function
is determined by the Ornstein-Zernike equation (\ref{eq:OZ}) and
the additional closure relation
\begin{equation}
c({\uu_1,\uu_2,\rr_{12}}) 
= (1- e^{V_{12}/k_B T}) \: (h(\uu_1,\uu_2,\rr_{12}) + 1),
\end{equation}
where $V_{12}$ is the pair potential between the particles 1 and 2.
In a homogeneous isotropic fluid, this defines a closed set of equations.
In an orientationally fluid, one needs an additional condition 
for the one particle distribution $\rho^{(1)}(\uu)$, e. g., eqn. 
(\ref{eq:gubbins}).

We have calculated the DCF in Percus-Yevick approximation for the
density values of our simulation. We followed the procedure described by 
Letz and Latz~\cite{letz}, except that we imposed eq. (\ref{eq:gubbins})
instead of isotropy. Nevertheless, we only obtained isotropic solutions. 
This is consistent with an argument of Zhong and Petschek, who showed
that the diagrammatic expansion of the Ward identity, an equation related 
to (\ref{eq:gubbins}), must contain diagrams in the anisotropic case 
which are excluded by the Percus-Yevick closure~\cite{zhong1}. 
Thus the Percus Yevick-closure is probably not compatible with 
spontaneous orientational ordering.

In the Percus-Yevick calculations, the spherical harmonics expansion 
was extended up to $l_{\mbox{\tiny max}}=4$. 
In addition, we have also performed systematic calculations with 
$l_{\mbox{\tiny max}}=2$. Here we observed an interesting scenario:
Over a wide range of densities, two solutions exist that meet in a
bifurcation at an upper critical density $\rho_c$ and disappear beyond
$\rho_c$. One of the branches merges into the Onsager solution at low 
densities. The other yields much more structured correlation functions 
and is clearly unphysical. Unfortunately, we have not been able to 
analyze systematically the situation at higher cutoff 
$l_{\mbox{\tiny max}}$. Our results suggest that the scenario at 
$l_{\mbox{\tiny max}} = 4$ might be similar. (For example, we found two 
solutions at $\rho = 0.29 \sigma_0^{-3}$). Therefore it 
may well be that the behavior observed at $l_{\mbox{\tiny max}}=2$ is a
generic-feature of the Percus Yevick approximation and not an artefact
of the truncation of $l$. This would explain how the Percus-Yevick solution 
can disappear at high densities without exhibiting a nematic 
instability~\cite{chamoux}.

Figure \ref{fig:py} compares the Percus-Yevick prediction for the DCF 
coefficient with $l_i=2,l=m=m_i=0$ with the simulation data for densities 
$\rho=0.24 \sigma^{-3}$ and $\rho=0.283 \sigma^{-3}$. 
The agreement is good even for the higher density, which is just below 
the transition. Above the transition, we can contrast the DCF with
the Percus-Yevick solution for the isotropic fluid of the same density.
For example, the Percus-Yevick DCF for $\rho = 0.3 \sigma_0^{-3}$, 
shown in figure \ref{fig:py} b) , can be compared with the real DCF, 
shown in figure \ref{fig:nisym}. The agreement is not very good.
The real, nematic DCF exhibits much more structure than the Percus-Yevick
result.

\fig{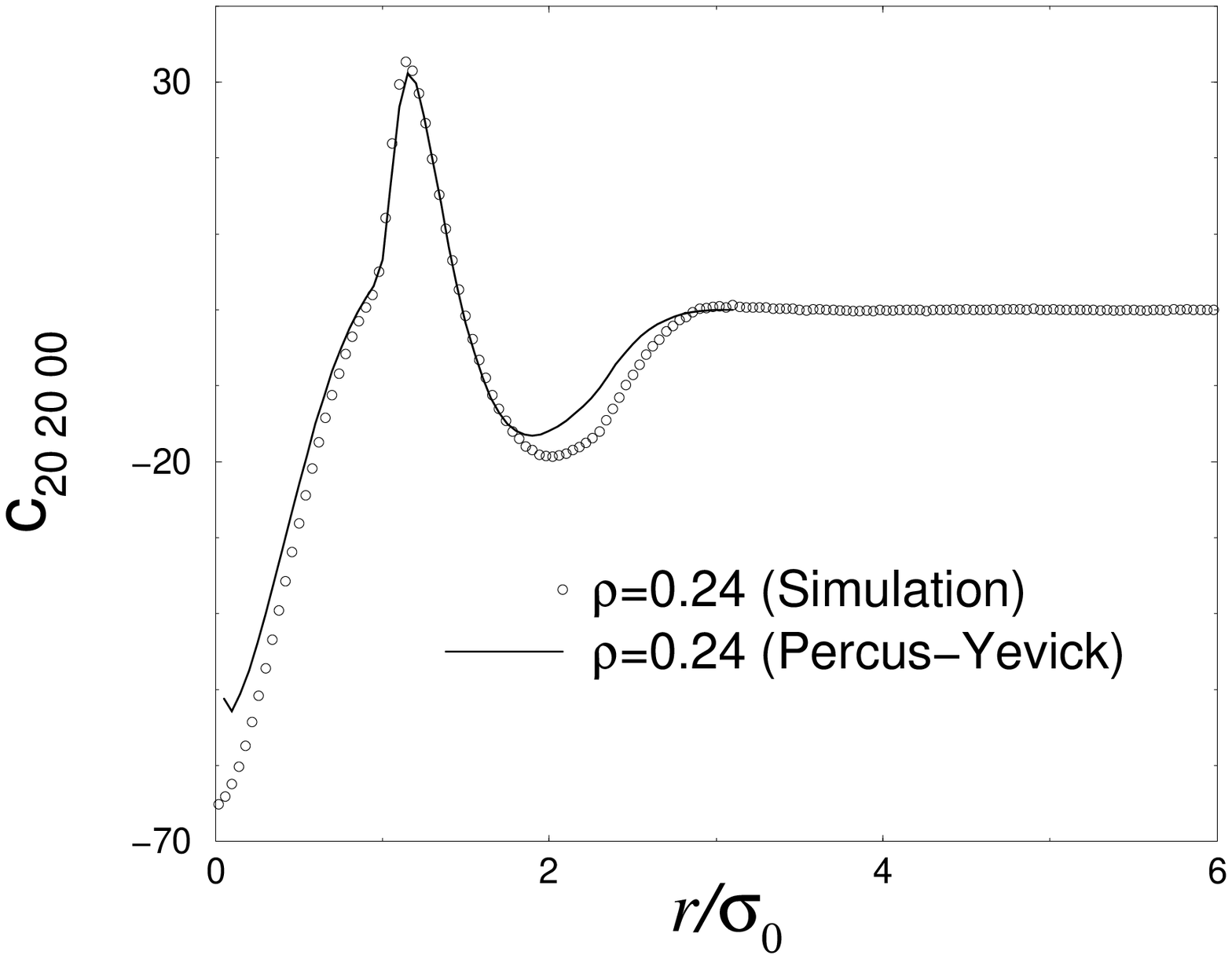}{70}{70}{(a)}

\fig{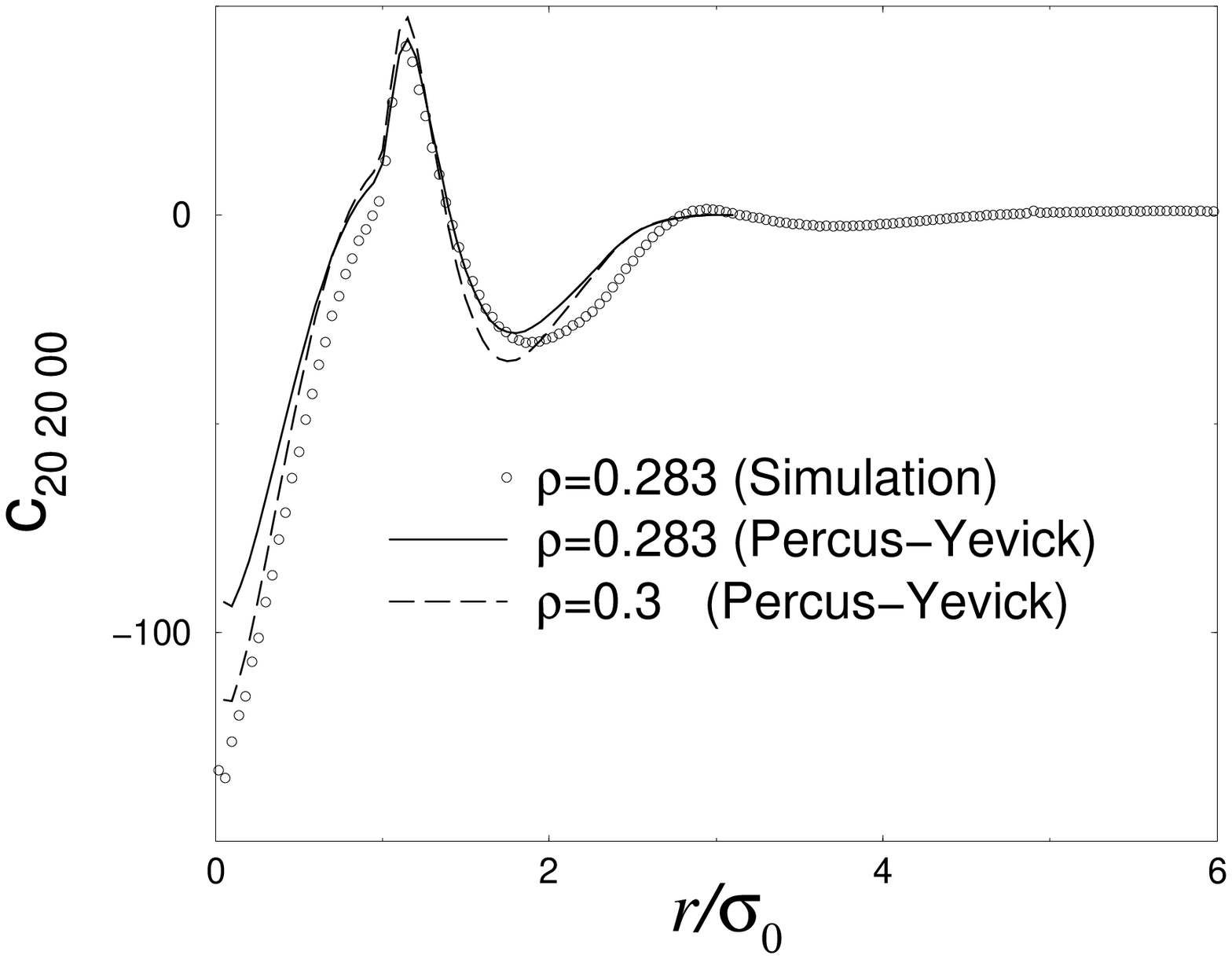}{70}{60}{(b)}

\begin{figure}
\CCpy
\end{figure}

\section{Summary and Conclusions}
\label{sec:summary}

To summarize, we have studied the local structure of fluids of uniaxial
ellipsoids at different densities in the vicinity of the nematic/isotropic
phase transition. In particular, we have calculated and analyzed the DCF.
We find that the DCF suitably characterizes the local structure, in the
sense that it is short ranged both in the isotropic and the nematic
phase. This reflects the short range nature of the interactions in our 
model. In other respect, however, the DCF does not reproduce the properties 
of the pair interaction potential. Noteably, it loses rotational invariance 
in the nematic phase and features a substantial symmetry breaking 
contribution. In contrast, the symmetry preserving part of the DCF does 
not change dramatically at the nematic/isotropic transition.

Our results should be useful guides for future developments of
density functional theories for anisotropic fluids~\cite{giorgio}, 
or of improved liquid state theories for nematic liquids in general. 
Moreover, we can use the DCF directly to predict the structure of 
inhomogeneous fluids of ellipsoids within simple density functional 
or integral equation approaches~\cite{hansen}. 
Such work is currently under way.

\section*{Acknowledgments}
 
We thank M.~P.~Allen, G.~Germano, and H.~Lange for fruitful interactions.
Some of the simulation data analyzed here have already been discussed in 
Ref.~\cite{phuong1}. These were produced in part by G.~Germano, using a
parallel MD program \textsc{gbmega} originally developed by the 
EPSRC Complex Fluids Consortium, UK. Both for the simulations and
the analyses, we have made extensive use of the CRAY T3E of the HLRZ in 
J\"ulich. This work was supported from the German Science Foundation (DFG).

\section*{Appendix}

In this appendix, we derive an equation which allows to convert
expansion coefficients of pair correlation functions from the
director frame to the molecular frame.

We denote by $F^{dir}(\uu_1,\uu_2,\rr)$ and $F^{mol}(\uu'_1,\uu'_2,\rr)$ 
the representation of a correlation function $F$ in the
director frame and the molecular frame, respectively.
These are expanded in spherical harmonics according to eqn. (\ref{eq:Fl}), 
hence 
\begin{eqnarray}
F^{dir}(\uu_1,\uu_2,\rr)& =& \! \! \!
\sum_{{{l_1,l_2,l}\atop {m_1,m_2,m}}} \!\!
F^{dir}_{l_1 m_1 l_2 m_2 l m}(r) \:
\\
&& 
Y_{l_{1}m_{1}}(\uu_1) \: Y_{l_{2}m_{2}} (\uu_2) \: Y_{lm}(\ru),
\nonumber
\end{eqnarray}
and
\begin{eqnarray}
\label{eq:emf}
F^{mol}_{l'_1 m'_1 l'_2 m'_2 l' m'}(r) & = &
\int d\uu_r d\uu'_1 d\uu'_2 
F^{mol}(\uu'_1,\uu'_2,\rr) \nonumber\\
&& 
Y^{*}_{l'_1m'_1}(\uu'_1)Y^{*}_{l'_2m'_2}(\uu'_2) Y^{*}_{l'm'}(\uu_r).
\end{eqnarray}
Now let ${\cal{R}}(\Omega) \equiv {\cal{R}}(\theta,\phi,\chi)$ be the rotation 
operator which carries the director frame into coincidence with the molecular 
frame (see figure \ref{fig:frames}), 
\begin{equation}
\label{eq:dm}
F^{mol}(\uu'_1,\uu'_2,\rr) = F^{dir}({{\cal{R}}\uu'_1},{{\cal{R}}\uu'_2},\rr).
\end{equation}
The angles $\theta$ and $\phi$ are the polar coordinates of the 
intermolecular distance vector $\rr_{ij}$ in the director frame,
and $\chi = \pi$. The coefficients of $F$ in the director frame 
and in the molecular frame are then related to each other by
\begin{equation}
\label{eq:emf1}
F^{mol}_{l'_1 m'_1 l'_2 m'_2 l' m'}(r) = \!\!\!
\sum_{{{l_1,l_2,l}\atop {m_1,m_2,m}}} 
K_{l_1m_1l_2m_2lm}^{l'_1m'_1l'_2m'_2l'm'}
F^{dir}_{l_1 m_1 l_2 m_2 l m}(r),
\end{equation}
where $K_{l_1m_1l_2m_2lm}^{l'_1m'_1l'_2m'_2l'm'}$ is given by
\begin{eqnarray}
\label{eq:K}
\lefteqn{K_{l_1m_1l_2m_2lm}^{l'_1m'_1l'_2m'_2l'm'} 
= \int d\uu_r d\uu'_1 d\uu'_2
Y^{*}_{l'_1m'_1}(\uu'_1) 
Y^{*}_{l'_2m'_2}(\uu'_2) 
} \qquad \qquad \qquad \\
&& \times 
Y^{*}_{l'm'}(\uu_r) 
Y_{l_1m_1}({\cal{R}}\uu'_1)Y_{l_2m_2}({\cal{R}}\uu'_2)Y_{lm}(\uu_r).
\nonumber
\end{eqnarray}
Next we apply the transformation formula~\cite{gray}
\begin{equation}
\label{eq:rmatrix}
Y_{l_1m_1}({\cal{R}}\uu'_1) = \sum_{M}D_{m_1M}^{l_1}(\Omega)^*
Y_{l_1M}(\uu'_1),
\end{equation}
where $D_{m_1M}^{l_1}(\Omega)$ is the rotation matrix of the operator 
${\cal{R}}$, given by
\begin{eqnarray}
\label{eq:D} 
D_{m n}^{l}(\theta,\phi,\chi)&=&e^{-im \phi}\sum_{k} (-1)^k
(\frac{1}{2})^l C_{lmnk} \\
&\times&
(1+a)^{l-k + \frac{m-n}{2}}
(1-a)^{k- \frac{m-n}{2}}e^{-i n \chi}
\nonumber
\end{eqnarray}
with $a = \cos\theta$ and 
\begin{equation}
C_{lmnk}
=\frac{\sqrt{(l+m)!(l-m)!(l+n)!(l-n)!}}{(l+m-k)!(l-n-k)!k!(k-m+n)!}.
\end{equation}
The sum over $k$ is taken over values such that the arguments of the  
factorials in the denominator of $C_{lmnk}$ are positive, i.e., 
max$(0,m-n) \leq k \leq$ min$(l-n,l+m)$.
Eq.~\ref{eq:K} then becomes 
\begin{eqnarray}
\label{eq:K1}
K_{l_1m_1l_2m_2lm}^{l'_1m'_1l'_2m'_2l'm'} &=& \int
d \uu_r Y^{*}_{l'm'}(\uu_r)Y_{l m}(\uu_r) \\
&&
D_{m_1 m'_1}^{l_1}(\Omega)^*\delta_{l_1 l'_1}
D_{m_2 m'_2}^{l_2}(\Omega)^*\delta_{l_2l'_2}.
\nonumber
\end{eqnarray}  

Using eqn. (\ref{eq:D}) with $\chi = \pi$ and the relation
\begin{equation}
\label{eq:SP}
Y_{lm} = \sqrt{\frac{(2l+1)}{4\pi}\frac{(l-m)!}{(l+m)!}}e^{im\phi}P_{lm}(a),
\end{equation}  
($P_{lm}$ is the associated Legendre polynomial), we obtain
\begin{eqnarray}
\label{eq:K2}
\lefteqn{
K_{l_1m_1l_2m_2lm}^{l'_1m'_1l'_2m'_2l'm'} \!\! = \!
\delta_{l_1l'_1}\delta_{l_2l'_2}
\delta_{m_1 +m_2 +m, m'} 
(-1)^{m'_1+m'_2}} \qquad 
&& \\
& \times &({\frac{1}{2}})^{l_1+l_2}
\sqrt{(2l+1)(2l'+1)}\sqrt{\frac{(l-m)!(l'-m')!}{(l+m)!(l'+m')!}}
\nonumber\\
&\times& 
\sum_{k_1,k_2}(-1)^{k_1+k_2}
C_{l_1m_1m'_1k_1} C_{l_2m_2m'_2k_2}\nonumber \\
&\times &\frac{1}{2}\int_{-1}^{1} da \: 
P_{l'm'}(a) \: P_{lm}(a) \:
(1+a)^{l_1 + l_2} \nonumber\\
&\times&
\sqrt{(1-a)/(1+a)}^{(- m'_1 - m'_2 +m_1 + m_2 + 2k_1 + 2k_2)}.
\nonumber 
\end{eqnarray}
In a uniaxial nematic phase, molecular frame coefficients 
with $m' \ne 0$ must vanish. Thus eqn. (\ref{eq:K2}) reduces to
\begin{eqnarray}   
\label{eq:Knematic}
\lefteqn{K_{l_1m_1l_2m_2lm}^{l'_1m'_1l'_2m'_2l'm'} = 
\delta_{l_1l'_1}\delta_{l_2l'_2} \delta_{m_1 +m_2+m,0} } \qquad &\\
&\times &({\frac{1}{2}})^{l_1+l_2}
\sqrt{(2l+1)(2l'+1)}\sqrt{\frac{(l-m)!}{(l+m)!}}\nonumber\\
&\times& \sum_{k_1,k_2}(-1)^{k_1+k_2}
C_{l_1m'_1m_1k_1} C_{l_2m'_2m_2k_2}\nonumber \\
&\times &\frac{1}{2}\int_{-1}^{1} \! \! da 
\: P_{l'}(a) \: P_{lm}(a) \:
(1+a)^{l_1 + l_2}. \nonumber \\
&\times& \sqrt{(1-a)/(1+a)}^{2k_1 + 2k_2 + m'_1 + m'_2} 
(-1)^{m'_1 + m'_2}  \nonumber
\end{eqnarray}

In an isotropic phase, one can use this result with 
$l'= m'=0$ and $m'_1 = -m'_2 = {\bar m}$.

The matrix for the inverse transformation from the molecular frame 
to the director frame the same, eqn. (\ref{eq:Knematic}).
 We have performed several numerical tests
to check this result: Pair distribution functions were
determined from the simulation data both in the molecular 
frame and in the director frame. Then we applied the
transformation formula and checked that the result
is the same~\cite{thesis}. In another test, we have 
transformed correlation functions into the molecular frame 
and back again, and compared the result with the original data.


\begin{thebibliography}{99}
\bibitem{footnote1} In the isotropic phase, a ``director''
  obviously does not exist. The choice of the coordinate
  system is then arbitrary. For reasons of consistency, 
  we still use the Eigenvector corresponding to the (small)
  largest Eigenvalue of the order tensor $\QQ$.
\bibitem{footnote2} Some restrictions still remain in uniaxial
  nematic fluids: All molecular frame coefficients are real
  and vanish unless $m=0$ and $l_1+l_2$ is even. In our case,
  $l_1$ and $l_2$ must also be even individually because of
  the symmetry of the particles.
\end{thebibliography}
\end{document}